\documentclass[conference]{IEEEtran}

\makeatletter
\def\ps@headings{%
\def\@oddhead{\mbox{}\scriptsize\rightmark \hfil \thepage}%
\def\@evenhead{\scriptsize\thepage \hfil \leftmark\mbox{}}%
\def\@oddfoot{}%
\def\@evenfoot{}}
\makeatother

\usepackage{times}
\usepackage{graphicx}
\usepackage{endnotes,url}
\usepackage[sort]{cite}
\usepackage{subfigure}
\usepackage{calc}
\usepackage{verbatim,amsfonts,amssymb}
\usepackage{amsmath}
\usepackage{multirow}
\usepackage{epsfig}

\newcommand{\ignore}[1]{}
\newcommand{\spacesave}[1]{}

\newcommand{\social}{SocialFilter}

\begin{document}

\title{SocialFilter: Collaborative Spam Mitigation Using Social Networks}

\author{\centering \rm
\begin{tabular}{ccc}
Michael Sirivianos & Xiaowei Yang & Kyungbaek Kim\\
Duke University & Duke University & University of California, Irvine\\
msirivia@cs.duke.edu & xwy@cs.duke.edu & kyungbak@uci.edu
\end{tabular}}

\maketitle

\begin{abstract}
Spam mitigation can be broadly classified into two main
approaches: a) centralized security infrastructures that rely on a
limited number of trusted monitors to detect and report malicious
traffic; and b) highly distributed systems that leverage the
experiences of multiple nodes within distinct trust domains. The first
approach offers limited threat coverage and slow response times, and
it is often proprietary. The second approach is not widely adopted, partly due to the 
lack of guarantees regarding the trustworthiness of nodes that comprise the
system.

Our proposal, \social, aims to achieve the trustworthiness of
centralized security services and the wide coverage, responsiveness
and inexpensiveness of large-scale collaborative spam mitigation. We propose a
large-scale distributed system that enables clients
with no email classification functionality to query 
the network on the behavior of a host. 
A \social\ node builds trust for its peers by auditing their behavioral 
reports and by leveraging the social network of \social\ administrators. 
The node combines the confidence its peers have in their own reports and the 
trust it places on its peers to derive the likelihood that a host is spamming.

The simulation-based evaluation of our approach indicates its
potential under a real-world deployment: during a simulated spam
campaign, \social nodes characterized $92\%$ of spam bot connections
with confidence greater than $50\%$, while yielding no false positives.
\end{abstract}

\section{Introduction}

The majority of the currently deployed spam email mitigation
techniques rely on centralized infrastructures and place trust on a
small number of security authorities. For instance, email systems and
browsers rely heavily on a few centralized IP reputation
services (e.g., ~\cite{SpamHaus,Singaraju-LISA-07,TrustedSource}). 

Unfortunately, centralized services often maintain out-dated
blacklists~\cite{Ramachandran-CCS-07}, offering a rather
large window of opportunity to spammers. Moreover, the vantage points
of centralized services are limited in numbers, but attacks launched using
large botnets are becoming increasingly surreptitious. In those
attacks, one malicious host may attack multiple domains, each for a short period of time
~\cite{Ramachandran-SIGCOMM-06,Katti-IMC-05}, reducing the effectiveness of spam traffic 
detection with a small number of vantage points. Finally, the use of such
email blacklisting services requires subscribing for a nominal fee when the service
is proprietary (e.g., Cloudmark~\cite{CloudMark} or TrustedSource~\cite{TrustedSource}.)

Motivated by this problem, researchers have proposed collaborative 
peer-to-peer spam filtering platforms~\cite{Zhong-INFOCOM-08,Zhou-Middleware-03} 
to achieve rapid and reliable detection and suppression of unwanted traffic.
These early systems assumed compliant behavior from all participating reporters of spam, which is hardly true given 
the heterogeneity of the Internet and the fact that reporters may belong to distinct trust domains. Compromised hosts controlled by attackers may join the
system, polluting the detection mechanisms. In addition, honest reporters may become
compromised after they join the system.  

To this end, we propose a collaborative spam filtering system
(\social) that uses social trust embedded in Online Social Networks
(OSN) to evaluate the trustworthiness of spam reporters. 
It relies upon the observation
that adjacent users in a social network tend to trust each other more than 
random pairs of users in the network. \social\ aims at aggregating the experiences of multiple security
authorities, democratizing spam mitigation. It is a trust layer that exports the likelihood that a 
host is spamming. Thus it enables nodes with no spam detection capability to collect the experiences 
of nodes with such capability and use them to classify connection requests from unknown email servers.

Each \social\ node submits {\em spammer reports}, i.e. security alerts regarding
Internet hosts identified by their IP address to a centralized repository peers (Section~\ref{sec:design}). 
The goal of the system is to ensure that the spammer reports reach other \social\ nodes 
prior to spamming hosts contacting those nodes, and that the spammer
reports are sufficiently credible to warrant action by their receivers. Each 
node associates a trust score to its peers and uses this score to assess the
trustworthiness of the reports originated by them.

\social\ uses social trust to bootstrap
direct trust assessments, and then employs a lightweight reputation
system ~\cite{Hoffman-ACM-07,Marti-ComNet-06} to evaluate the trustworthiness of nodes 
and their spammer reports (Section~\ref{subsec:determining-reporter-trust}).
Our insight is that each node will be administered by human
administrators (admins), and nodes maintained by trusted admins are
likely to disseminate trustworthy reports. Therefore, a \social\ node
may obtain a direct trust assessment with a number of nodes with whom
its admin has social relationships.  Social relationships between
admins can be obtained from massive OSN providers, such as Facebook and LinkedIn. 

However, reputation systems are known to be vulnerable to the Sybil
attack~\cite{Douceur-IPTPS-02}. Sybil attacks subvert
distributed systems by introducing numerous malicious identities 
under the control of an adversary. By using these
identities the adversary acquires disproportional influence over 
the system. To mitigate this attack,
\social\ again uses the social network to assess the probability
that a node is a Sybil attacker, i.e. its  {\em identity uniqueness} (Section~\ref{subsec:osn-certification-authorities}).
Each node's identity is associated with its
admin's identity. The latter is verified through the social network
using a SybilLimit-like technique~\cite{Yu-SP-08}, which can effectively 
identify Sybils among social network users.

\social\ nodes use both the identity uniqueness and
the reputation of another node to assess the overall trustworthiness
of that node's report. The originator of a spammer
report itself also assigns a confidence level to the report, as
traffic classification has a level of uncertainty.  The
trustworthiness of a node and its confidence level in a report
determines whether a node should trust a report or ignore it. Trusted
reports can be used for diverse purposes, depending on the 
node's function. For example, email servers can use them to
automatically filter out email messages that originate from IPs that
have been designated as spammers with high confidence.  IDS systems
can use them to block SMTP packets that originate from suspicious IPs.

A recent unwanted traffic mitigation system, 0stra~\cite{Mislove-NSDI-08}, 
combat unwanted traffic by forcing it to traverse social links
the capacity of which imposes a rate-limit over the communication.
Unlike Ostra, \social\ does not use social links to
rate-limit unwanted traffic. Instead it utilizes social links
to bootstrap trust between nodes and to suppress Sybil attacks.
However, Ostra can result in legitimate email being blocked
(false positives), which is highly undesirable.

We evaluate our design (Section~\ref{sec:evaluation}) 
using a $50K$-node sample of the Facebook
social network.  We demonstrate through
simulation that collaborating \social\ nodes are able to suppress
common types of unwanted traffic in a reliable and responsive
manner. Our simulations show that in a $50K$-node \social\ network
with only $10\%$ of nodes having spam classification capability, 
nodes with no local spam detection capability are able to identify $92\%$ 
of connections from spammers with greater than $50\%$ confidence. 
Our experimental comparison with Ostra shows that our approach
is slightly less effective in suppressing  spam, while in contrast
to Ostra, it yields no false positives. Given the severity of the 
problem of false spam positives, we believe that \social\ can be a better alternative
under many scenarios. 

\section{System Overview}
\label{sec:system-overview}

In this section, we provide a high-level description of our system
and the security challenges it addresses.

\subsection{\social\ Components}
\label{subsec:socialfilter-components}

\begin{figure}[t]
\centering
\includegraphics[scale=0.3]{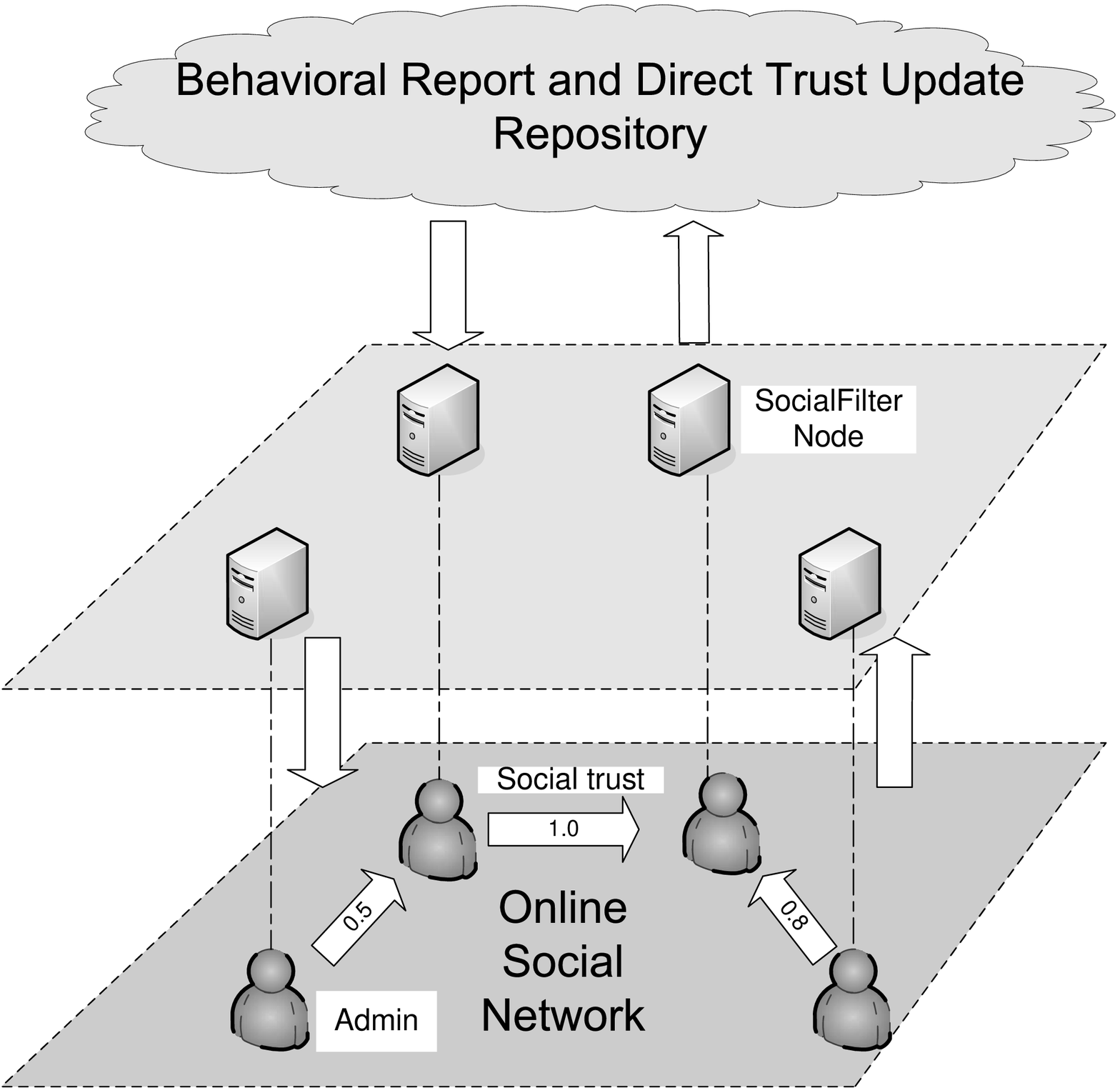}
\caption{\label{fig:socialfilter-architecture} \social\ architecture.}
\end{figure}

Figure~\ref{fig:socialfilter-architecture} depicts \social's architecture.  
At a high-level, the \social\ system comprises the following components: 1)~human
users that administer networked devices/networks
({\em admins}) and join a social network; 2)~end systems  ({\em \social\ nodes}) that are
administered by specific admins and participate in monitoring and
reporting the behavior of email senders;
3){\em~spammer reports} submitted by \social\ nodes concerning email senders
they observe; 4)~direct trust updates made available by \social\ nodes reporting
their perceived trustworthiness of their peers; and 5)~a centralized
repository that receives and stores spammer reports and trust
updates. 

The same admin that administers a \social\ node also administers
a group of applications that interface with the node to report spamming behavior.
Interfacing applications can either be SMTP servers or IDS systems~\cite{Paxson-ComNet-99}
that register with the \social\ repository or users of a webmail service. In the first case SMTP servers can classify 
spam by using their email characterization functionality (host reputation services such as TrustedSource~\cite{TrustedSource}, 
CloudMark~\cite{CloudMark} and DShield~\cite{Zhang-Security-08}, or content-based filters).
In the second case, the interfacing application is essentially a human user who reports
an email (and consequently its originating email server) as spam.

\subsection{Spammer Reports}
\label{subsec:spammer-reports}

An email characterization application uses the {\tt ReportSpammer(hostIP, confidence)}
call of the \social\ node RPC API to feedback its observed behavior for an email sender. 
The first argument identifies the source, i.e., an IP address. The second 
argument is the confidence with which the application is reporting that the specified
entity is a spammer. The latter takes values in $0\%$ to $100\%$
and reflects the fact that in many occasions traffic classification 
has a level of uncertainty. For example, a mail server that sends both spam and legitimate
email may or may not be a spamming host.

In turn, the \social\ node submits a corresponding spammer
report to the repository to share its experience with its peers. 
For example, if a node $i$'s spam analysis indicates that half of the emails
received from host with IP $I$ are spam, $i$ reports:
\begin{center} ${[\rm spammer~report]}~I,~50\%$ \end{center}
In addition, \social\ nodes are able to revoke spammer reports by updating
them. If for example at a latter time, $i$ determines that no spam originates from $I$, it 
sends a new report in which it updates the confidence value from $50\%$ to $0\%$.
Nodes can authenticate with both the repository and the OSN provider using standard single-sign-on
authentication techniques, e.g., \cite{shibboleth,Steiner-USENIX-88} or Facebook Connect~\cite{facebook-connect}.

\subsection{Determining whether a Host is Spamming}
\label{subsec:trusting-a-host}

Each node assigns a {\em peer trust} value
to each of its peers in the network. This trust value 
determines the trustworthiness of the peer's spammer reports.
The {\em peer trust} is determined using two dimensions of trust: a) {\em reporter trust}; 
and b) {\em identity uniqueness}. We describe these two dimensions below.
The receiver of a spammer report derives its confidence in the 
correctness of the received report from the report's confidence
and the peer trust. 

Nodes collectively compute {\em reporter trust} values
by employing a  reputation management mechanism. 
This mechanism relies on \social\ nodes verifying each other's 
spammer reports to derive individual {\em direct trust} values (Section~\ref{subsec:determining-reporter-trust}).
If a node $i$ is able to verify the spammer reports of node $j$ 
it can determine a direct trust value $d_{ij}$.
The \social\ nodes share these values with other nodes by exchanging
{\em direct trust updates}. For reasons of scalability and efficiency, a node $i$
considers the spammer reports of only a (possibly random) subset $V_i$ 
(including itself) of all the nodes in the \social\ network. 
Consequently, nodes submit and retrieve direct trust updates 
only for nodes in $V_i$. We refer to $V_i$ as node's $i$ {\em view}.

Our system also relies on the fact that nodes comprising Internet systems such as 
email servers, honeypots, IDS etc are administered by human admins. 
These human users maintain accounts in online social networks (OSN). \social\
leverages  OSNs in the following two ways: a) it defends against Sybil attacks~\cite{Douceur-IPTPS-02} 
by exploiting the fact that OSNs can be used for resource testing, where the test in question is a 
Sybil attacker's ability to create and sustain acquaintances. Depending on the result of the test,
the OSN provider assigns an {\em identity uniqueness} value to the admin;
and b) It initializes the {\em direct trust} values in the absence 
of prior interactions between \social\ nodes, by considering the trust that is inferred by associations 
in the social network of administrators. 

Finally, an application can use the {\tt IsSpammer(hostIP)} call
of the \social\ node RPC API to obtain a value that corresponds
to the likelihood of the hostIP being spamming. The node 
derives this value by aggregating spammings reports
regarding hostIP. These reports are weighted
by the reporting node's {\em peer trust}. 


\subsection{Security Challenges}
\label{subsec:security-challenges}

\social\ is a collaborative platform aiming at suppressing
malicious traffic. In addition, it is an open system, 
meaning that any admin with a social network account and a device can join.
As such, it is reasonable to assume that \social\ itself will be 
targeted to  disrupt its operation. Our system faces
the following security challenges:

\noindent \textbf{False Spammer Reports.} Malicious 
\social\ nodes may issue false or forged reports aiming at 
reducing the system's ability to detect spam or disrupting
legitimate email traffic.

\noindent \textbf{False Direct Trust Updates.} To address false 
spammer reports, \social employs a reporter reputation system to 
determine the amount of trust that should be placed on each user's
reports. However, the reputation system itself is vulnerable to false 
reporting as malicious nodes may send false or forged direct trust updates.

\noindent \textbf{Sybil Attack. } An adversary may attempt to create
multiple \social\ identities aiming at increasing its ability to subvert
the system using false spammer reports and direct trust updates.
Defending against Sybil attacks without a trusted central authority is hard.  
Many decentralized systems try to cope with Sybil attacks by binding an
identity to an IP address. However, malicious users can readily harvest
IP addresses through BGP hijacking~\cite{Ramachandran-SIGCOMM-06} or 
by commanding a large botnet.

\section{\social\ Design}
\label{sec:design}

We now present in  more detail the design of our system. 

\subsection{OSN Providers as Sybil Mitigating Authorities}
\label{subsec:osn-certification-authorities}

For an open system such as \social\ to operate reliably, 
node accountability in the form of node authentication 
and prevention of Sybil attacks is of the utmost importance.

We propose to leverage existing OSN repositories as inexpensive, Sybil-mitigating 
authorities. OSNs are ideally positioned to perform such function: 
Using SybilLimit-like~\cite{Yu-SP-08} techniques (see Section~\ref{subsubsec:identity-uniqueness})
to perform inexpensive resource tests on the social graph, OSNs can determine the amount of
confidence one can place on a node's identity. We refer to this confidence as {\em identity uniqueness}.

Each node that participates in  \social\ is administered by human users that have 
accounts with OSN providers. The system needs to ensure that each user's social
network identity is closely coupled with its \social\ node. To this end, \social\  
single sign-on authentication mechanisms, such as Facebook Connect, to associate 
the OSN account with the spammer report and direct trust update repository account.

\subsubsection{Determining the Identity Uniqueness}
\label{subsubsec:identity-uniqueness}

When malicious users create numerous fake online personas, \social's
host trust metric can be subverted.  Specifically, a malicious user $a$ 
with high reporter trust with another user $u$ may create Sybils and 
assign high direct trust to them. As a result, all the Sybils
of the attacker would gain high reporter trust with user $u$. 

There is typically one-to-one correspondence between a real user's social network 
identity and its real identity. Although, malicious users can create many identities, they 
can establish only a limited number of trust relationships with real humans. Thus, groups 
of Sybil attackers are connected to the rest of the social graph with a disproportionally 
small number of edges. The first works to exploit this property was SybilGuard  and SybilLimit
\cite{Yu-SIGCOMM-06,Yu-SP-08}, which bound the number of Sybil identities using a 
fully distributed protocol.

Based on a similar concept, \social's  Sybil detection algorithm determines the uniqueness
of a \social\ user's identity. This algorithm is executed solely by the OSN provider over its
centrally maintained social graph. An admin's $i$ identity is considered weak if it 
has not established a sufficient number of real relationship's over the social network. Upon being 
queried by an admin $v$, the OSN provider returns a value in $[0.0,1.0]$, which specifies the confidence 
of the provider that a specific node $s$ is not participating in a Sybil attack, i.e. the probability
that $s$ is not part of a network of Sybils.

First, we provide some informal background on the theoretical justification of 
SybilGuard and SybilLimit. It is known that randomly-grown topologies such as social
networks and the web are fast mixing small-world topologies~\cite{Watts-Nature-1998,Albert-Nature-1999,Kaiser-Science-1999}.
Thus in the social graph $\mathcal{I}$ with $n$ nodes, a walk of $\Theta(\sqrt{n} \log n)$ steps 
contains $\Theta(\sqrt{n})$ independent samples approximately drawn from the 
stationary distribution. When we draw random walks from a verifier node $v$
and the suspect $s$, if these walks remain in a region of the network 
that honest nodes reside, both walks draw $\Theta(\sqrt{n})$ independent samples from 
roughly the same  distribution. It follows from the generalized Birthday Paradox~\cite{Yu-SIGCOMM-06}
that they intersect with high probability. The opposite holds if the suspect resides
in a region of Sybil attackers that is not well-connected to the region of honest
nodes. 

SybilGuard replaces random walks with ``random routes'' and a verifier
node accepts the suspect if random routes originating from both nodes 
intersect. In random routes, each node uses a pre-computed random 
permutation as a one-to-one mapping from incoming edges to outgoing 
edges. Each random permutation generates a unique routing table 
at each node. As a result, two random routes entering an honest
node along the same edge will always exit along the same edge (``convergence property''). 
This property guarantees that random routes from a Sybil region that is connected to the honest
region through a single edge will traverse only one distinct path, further reducing the 
probability that a Sybil's random routes will intersect with a verifier's random routes.
SybilLimit~\cite{Yu-SP-08} is a near-optimal improvement over the SybilGuard
algorithm. In SybilLimit, a node accepts another node only if random routes
originating from both nodes intersect at their last edge.
For two honest nodes to have at least one intersected last edge with high probability,
the required number of the random routes from each node should be approximately
$r=\Theta(\sqrt{m})$, where $m$ is the number of edges in $\mathcal{I}$.
The length of the random routes should be $w=O(\log n)$.

With \social's SybilLimit-like technique the OSN provider computes an identity uniqueness
score for each node $s$ in the social graph $\mathcal{I}$. At initialization time, the OSN provider 
selects $l$ random verifier nodes. It also creates $2r$  independent 
instances of pre-computed random permutation as a one-to-one mapping from 
incoming edges to outgoing edges (routing table). The first $r$ routing tables are used
to draw random routes from suspect nodes $s$ and the rest $r$ routing tables are used
to draw random routes from the verifier nodes $v$. SybilLimit uses distinct routing
tables for verifiers and suspects in order to avoid undesirable correlation between 
the verifiers' random routes and the suspects' random routes. 
For each $s$, the OSN provider runs the SybilLimit-like algorithm is as follows:

\begin{enumerate}
\item For each of the $l$ verifiers $v$, it picks a random neighbor
of $v$. It draws along the random neighbors $r$ random routes of length
$w=O(\log n)$, for each instance of the $r$ routing tables,
where $n$ is the number of nodes in $\mathcal{I}$. It stores the last edge
(tail) of each verifier random route.
\item It picks a random neighbor of $s$ and draws along it $r$ random
routes of length
$w=O(\log n)$, for each instance of the nodes' routing tables.
It stores the last edge (tail) of each suspect random route.
We refer to steps $(1)$ and $(2)$ of the algorithm as {\em random routing.}
\item For each verifier $v$, if one tail from $s$ intersects one tail from $v$,
that verifier $v$ is considered to ``accept'' $s$. We refer to
this step as {\em verification.}
\item It computes the ratio of the number of verifiers that accept $s$ over the
total number of verifiers $l$. That ratio is the computed identity uniqueness
score $id_s$.
\end{enumerate}

Nodes query the OSN provider for the identity uniqueness of their peers. The
OSN provider performs the above computations periodically and off-line to
accommodate for topology changes. The OSN provider stores the result of this 
computation for each node as a separate attribute.

\subsection{Determining the Reporter Trust}
\label{subsec:determining-reporter-trust}

Malicious nodes may issue false spammer reports to manipulate the trust towards entities.
In addition, misconfigured nodes may also issue erroneous reports.
\social\ can mitigate the negative impact of incorrect 
reports by assigning higher weights to reports obtained from more 
trustworthy \social\ nodes.

Conceptually, each \social\ node $i$ maintains a
reporter trust value $rt_{ij}$ to every other  node $j$ in its view, 
$j\in V_i\setminus i$.  This trust score corresponds to node $i$'s estimation of the probability
that node $j$'s reports are accurate. It is obtained from three
sources: trust attainable from online social networks, direct spammer
report verification, and transitive trust.

First, \social\ relies on the fact that \social\ nodes are
administered by human users. Competent and benign users are
likely to maintain their nodes secure, and provide honest and truthful
reports. The trust on the competency and honesty of human users could
be obtained via social networks.  \social\ admins
maintain accounts in online social networks. An admin $i$ tags her
acquaintance admin $j$ with a social trust score $st_{ij}$ in $[0.0,1.0]$ based
on her belief on $j$'s ability to manage her node(s).  This
value is used to initialize a direct trust score between two 
nodes $i$ and $j$: $d_{ij} = st_{ij}$. 

Second, a \social\ node $i$ dynamically updates the direct trust
$d_{ij}$ by comparing spammer reports submitted by the node $j$
with its own submitted reports. A node $i$ may verify a report from a node $j$
for an entity $e$, if $i$ has also recently interacted with the same entity. 
$i$ may also probabilistically choose to observe $e$ solely for the purpose 
of verifying reports of another node $j$. The portion of  the received spammer
reports that the \social\ nodes verify 
is a tunable parameter. Intuitively, if $i$
and $j$ share similar opinions on $e$, $i$ should have a high trust
in $j$'s reports. Let $v^k_{ij}$ be a measure of similarity in
$[0,~1.0]$ between $i$ and $j$'s $k_{th}$ report. A node $i$  updates its
direct trust to $j$ using an exponential moving average:
\begin{equation}
\label{eq:direct-trust}
d^{k+1}_{ij} = \alpha * d^{k}_{ij} + (1-\alpha) * v^{k+1}_{ij}
\end{equation}

As $i$ verifies a large number of reports from $j$, the direct trust metric
$d^k_{ij}$ gradually converges to the similarity of reports from $i$
and $j$. 

By updating $d_{ij}$ and making it available for retrieval to other nodes, $i$ enables 
its peers $j\in V_i$ to build their reporter trust graph $T_j(V_j,E_j)$. 
The reporter trust graph of a node $i$ consists of only the nodes in  its view $V_i$, and its directed 
edge set $E_i$ consists of the direct trust $d_{uv}$ for each $u,v\in V_i$.
If a node $u$ has not released a direct trust update for a node $v$, 
$d_{uv}$ is treated as being equal to $0.0$.

Third, a node $i$ incorporates direct trust and transitive
trust~\cite{Gray-LNCS-03,Guha-WWW-04} to obtain $i$'s overall trust to $j$: $rt_{ij}$.  We use
transitive trust for the following main reasons: a) due to the
large number of nodes, the admin of a  node $i$
may not assign a social trust $s_{ij}$ to the admin of a node $j$,
as they may not be acquainted;  b) due to the large number of email-sending hosts, 
nodes $i$ and $j$ may not have encountered the same hosts and are
therefore unable to directly verify each other's reports; and 
c)  $i$ can further improve the accuracy of its trust
metric for $j$ by learning the opinions of other \social\ nodes
about $j$. The overall reporter trust $rt_{ij}$ can be obtained as the maximum
trust path in node $i$'s reporter trust graph $T_i(V_i,E_i)$, 
in which each edge $u \rightarrow v$ is annotated by the direct trust $d_{uv}$. That is,
for each path $p \in P$, where $P$ is the set of all paths between nodes $i$ and $j$:
\begin{equation}
\label{eq:reporter-trust}
rt_{ij} = max_{p\in P} (\Pi_{u \rightarrow v \in p} d_{uv}) 
\end{equation}

We use the maximum trust path because it can be efficiently computed
with Dijkstra's shortest path algorithm in $O(|E| \log |V|)$ time for a sparse $T$. 
In addition, it yields larger trust values than the minimum or average trust path, resulting
in faster convergence to high confidence regarding the actions entities perform. 
Finally, it mitigates the effect of misbehaving nodes under-reporting their trust 
towards honest nodes.  Messages that appear spamming to a node 
may not appear so to all other nodes in the system. For example a compromised host
may send spam to certain hosts, but at the same time may send legitimate emails to others.
Therefore the subjective local trust metric we use is more appropriate than a global trust 
metric, such as Eigentrust's~\cite{Kamvar-WWW-03}. 

The false direct trust update attack mentioned in Section~\ref{subsec:security-challenges} 
may manifest in two ways. First, a misbehaving reporter $x$ in a node's $w$
view sends false direct trust updates regarding another node $y$
in $w$'s view. Second, a source of spam $s$  sends  good traffic to node $x$ and spam
traffic to node $y$, while node $x$ verifies $y$'s reports.
As a result $x$ will perceive $y$ as not being trustworthy. Thus, a
node $w$ that has both $x$ and $y$ in its view would 
incur the false direct trust update attack.
If for example node $w$ trusts $x$ by $1.0$ and node $x$ trusts $y$ by $0.0$, 
$w$ would trust $y$ by 0.0 and would no longer consider its reports valid, 
although $y$'s reports are correct.
However, our design is inherently resilient to this attack as we
demonstrate in Section~\ref{subsec:socialite-resilience-to-attacks} (Figure~\ref{fig:VariousFalseReporters}):
if the node $w$ has many neighbors and possibly alternative trust paths to $y$
or receives spammer reports from a large number of nodes in its view, 
this attack is mitigated. Also this attack would have an effect only 
against nodes that have both $x$ and $y$ in their view. In addition, 
the attacker should have a legitimate reason to send traffic to $x$.

\subsection{Determining the Likelihood of a Host being Spammer}
\label{subsec:determining-entity-trust}

As mentioned above, a \social\ node $i$ may receive multiple spammer reports
originating from multiple nodes $j\in V_i$ and concerning the same host $h$
for the same action $a$. Each report concerning $h$ is marked with the  level of 
confidence $c_j(h)$ of the reporter $j$. For example, this confidence may
be equal to the portion of emails received by  host $h$ that are spam (~\cite{Singaraju-LISA-07}).
Subsequently, $i$ needs to aggregate the spammer reports to determine
an overall likelihood {\tt IsSpammer($h$} that $h$ is a spam bot.

When a node $i$ that does not have entity classification functionality
receives multiple reports concerning the same host $h$, it derives the overall 
likelihood {\tt IsSpammer($h$)} weighing the spammer reports' confidence by the 
peer trust of their reporters:
\begin{equation}
\label{eq:spammer-report-trust}
IsSpammer(h) = \frac{\Sigma_{j \in V_i\setminus i}~rt_{ij}~id_j~c_j(h)} {\Sigma_{j \in V_i\setminus i}~rt_{ij}~id_j}
\end{equation}

If applications interfacing with node $i$ have entity classification functionality,
and sent to $i$ spammer reports through the {\tt ReportSpammer()} interface, 
$i$ considers only these reports in calculating the 
trust for an entity.  When $i$ receives spammer reports by more than one applications
for the same $h$, the confidence that the node has in $h$ is the 
average (possibly weighted) of these applications' reports. Node $i$ uses this average confidence
to compute the similarity of its reports with the reports its peers.
When a node $i$ receives a new spammer report for $h$, this new report preempts an 
older report, which is thereafter ignored.

Each spammer report carries a timestamp. The time interval $T$ during which
a spammer report is valid  is a tunable system parameter. Reports for which 
$current-time - timestamp > T$ are not considered in the calculation of the 
likelihood of a host being spamming. We assume lose synchronization between 
\social\ nodes.

\subsection{SocialFilter Repository}
\label{subsec:socialfilter-repository}

\begin{figure*}[ht!]
\centering
\includegraphics[width=0.8\textwidth]{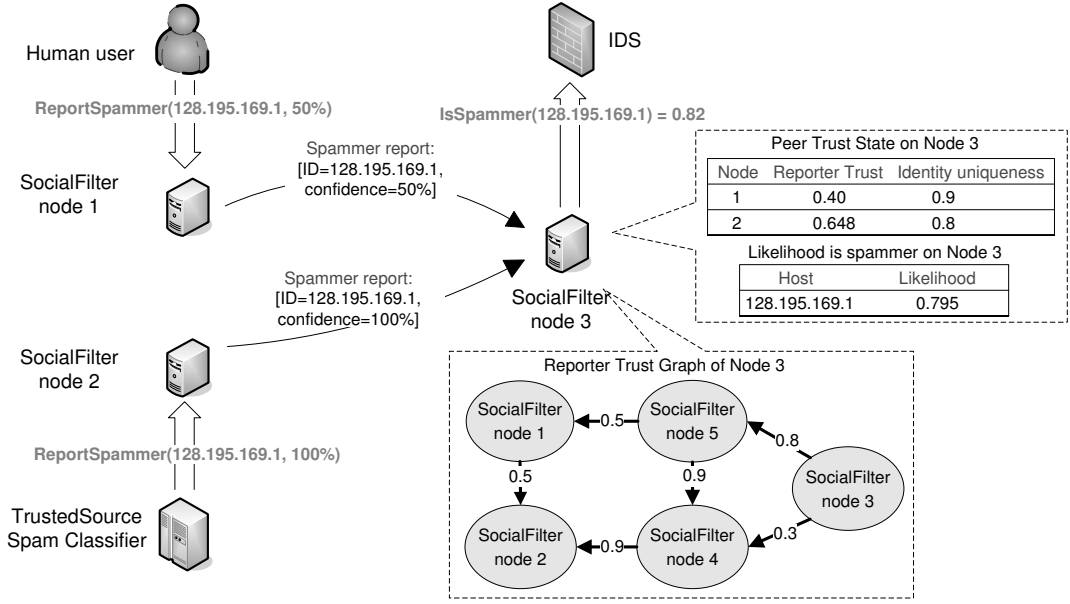}
\caption{\footnotesize{Example of the operation of a small \social\ network.}}
\label{fig:socialfilter-operation-example}
\end{figure*}

A node  can exchange spammer reports and direct trust
updates with any other node in the \social\ network regardless of 
whether the admins of the nodes are acquaintances in the social 
network. With this design choice, we ensure that spammer reports and 
direct trust updates reach the  interested nodes 
on time, improving the threat coverage of our system.
We also enable users that are not well-connected in the social network  
to peer with other trustworthy nodes.

Our centralized repository consists of two parts, one for  spammer
reports and one for direct trust updates. The portion of the repository 
tasked with maintaining spammer reports is implemented as 
a hash table. Nodes store and retrieve spammer reports
concerning nodes in their view. When a node queries for 
spammer reports, it is interested on the reports for a single entity/action 
pair. These reports are sent by multiple nodes, thus for efficiency
it is reasonable to index(key) them based on the hash of the 
concatenation of the entity's ID (e.g., IP) and the action description.

When a node $i$ encounters a specific entity, it queries the repository 
for all the spammer reports that involve the entity and the action. Once it locates
the node that stores those reports it asks the node for
those reports that originate from nodes in $V_i\setminus i$. 

On the other hand, a node needs to retrieve all the direct
trust updates involving all the nodes in its view. 
Thus, it is reasonable to also implement the direct trust update
repository as a hash table indexed by the ID of the node that issues the update. 
A node $i$ needs to explicitly query for the existence of an update
involving all node pairs in its view. Thus every time interval $D$, a node 
$i$ requests from the repository for each node $j\in V_i\setminus i$ for the 
current non-zero direct trust values $d_{jv}$ for $v\in V_i$. Using these
direct trust values $i$ can build the trust graph  of its view $V_i$ $T_i(V_i,E_i)$.
If the difference between the current direct trust metric $d_{jv}$
and the last $d_{jv}$ $i$ retrieved from repository is greater than $\epsilon$, 
the repository includes this update in his reply to $i$'s request for direct trust updates. 
The constant $\epsilon$ is used to ensure that the repository does not incur the 
overhead of communicating the update if it is not sufficiently large.

\subsection{\social\ Operation Example}
\label{subsec:socialfilter-operation-example}

Figure~\ref{fig:socialfilter-operation-example} depicts an example of the operation
of a small \social\ network. The network includes an IDS node tasked
with checking incoming TCP connections for whether they originate from spamming hosts, \social\ node 3. That node
has no inherent email classification functionality, thus it relies on the other
two nodes, 1 and 2, for early warning about spam bots. Node 1 relies on human users
to classify emails as spam. In this example, the human user has classified half
of the emails originating from host with IP=128.195.169.1 as spamming, therefore
it reports that this host is spamming with confidence $c_1(IP)=50\%$. Node 2 is an email
server that has subscription to a proprietary blacklisting service. In this example
Node 2 received a connection request from the host with IP=128.195.169.1, it queried the reputation service 
and got a response that this host is spamming. Thereby, node 2 reports 
with confidence $c_2(IP)=100\%$ confidence that the host is a spam bot.

Node 3 maintains the depicted reporter trust graph,
derived from its $5$-node view. This view includes nodes 1 and 2, which sent
the depicted spammer reports. It also includes nodes 4 and 5, which did not sent
any reports in this example. The weighted directed edges in the graph correspond to 
the direct trust between the peers in node 3's view.
From the reporter trust graph and Equation~\ref{eq:reporter-trust}, 
the maximum trust path between nodes 3 and 1 traverses nodes 5 and 1 yielding $rt_{31}=0.4$.
The maximum trust path between 3 and 2 traverses nodes 5, 4 and 2 and yields 
reporter trust $rt_{32}=0.648$. The identity uniqueness
of nodes 1 and 2 has been computed by the OSN provider to be $id_1=0.9$ and $id_2=0.8$,
respectively. We can now use Equation~\ref{eq:spammer-report-trust} to 
compute the confidence {\tt IsSpammer(IP)} that the IDS interfacing with node 3 
has that the host is spamming:
\[\frac{rt_{31} id_1 c_1(IP) + rt_{32} id_2 c_2(IP)} {rt_{31} id_1
+ rt_{32} id_2} = 0.795\]

\section{Evaluation}
\label{sec:evaluation}

We evaluate \social's ability to block spam traffic and compare it to Ostra~\cite{Mislove-NSDI-08}. 
Ostra represents a different approach to spam mitigation using social links, and the main goal
of this evaluation is to shed light on the benefits and drawbacks of the two approaches.

\subsection{Ostra Primer}
\label{subsec:ostra-primer}

Before we proceed with the comparative evaluation, we first provide an overview of Ostra.
Ostra bounds the total amount of unwanted communication a user can send based on the number
of trust relationships the user has and the amount of communication that has been 
flagged as wanted by its receivers. Similar to \social\ in Ostra, an 
OSN repository maintains the social network. When a sender wishes to send email to a  receiver, 
it first has to obtain a cryptographic token from the OSN repository. The OSN
repository uses the credit balances along the social links connecting the sender
and the receiver to determine whether a token can be issued. Each user adjacent
to a social link is assigned a credit balance, $B$, which is unique for the link. 
$B$  has an initial value of 0. Ostra also maintains a per-link balance
range $[L,U]$, with $L$ $\leq$ 0 $\leq$ $U$, which limits the range of
the user’s credit balance (i.e., always $L$ $\leq$ $B$ $\leq$ $U$).
The balance and balance range for a user is denoted as $B^U_L$. 
For instance, the link's adjacent user's state $2^{+5}_{-4}$ denotes that 
the user's current credit balance is $2$, and it can range between $-4$
and $5$.

When a communication token is issued, Ostra requires that there is
a path between the sender and the receiver in the social network. It then
requires that for each link along the social path the first adjacent node’s 
credit limit $L$ is increased by one, and the second adjacent node’s 
credit limit $U$ is decreased by one. This process propagates recursively 
from the sender to the receiver along the social links.  If this process
results in any of the links in the path to have adjacent nodes of which the credit balances 
exceed the balance range, Ostra  refuses to issue the token. When the
communication is classified by the receiver, the credit limits $L$ and $U$
are restored to their previous state. If the communication is marked as unwanted, one credit  is transferred from the 
balance of the first node of the link to the balance of the second one. 
Eventually, the links that connect spammers to their receivers
have balance beyond the allowed range and  a spammer is prevented from sending email.

\subsection{Evaluation Settings}

For our evaluation, we use a large strongly connected component sampled from 
Facebook, consisting of 50,000 nodes and 442,772 symmetric links.

We use the SimPy 1.9.1~\cite{SimPy} simulation package to simulate SocialFilter and Ostra under a
scenario where the social network is formed among the admins of email servers.
We assume that legitimate users usually send 3  emails per day. 80\% and 13\% of the legitimate emails
are sent to sender's friends and sender's friends of friends respectively, and the destination of the rest 7\% 
emails are randomly chosen by the sender. There are some spammers, which also participate,
in the Ostra and \social\ network, sending 500 spam emails per day each to random users in the network.  
In this evaluation we set Ostra's credit bounds equal to 5 ($|L|=|U|=5$). These settings are obtain from 
Ostra's evaluation~\cite{Mislove-NSDI-08}.

Several nodes can automatically classify spam emails. These instant classifiers correspond to systems that detect spam 
by subscribing to commercial blacklists, employ content-based filters etc. These nodes  can block spam email instantly. 
On the other hand, normal users can classify an email only after receiving/reading it.
That is, the normal classification can be delayed based on the behavior of the users
(how frequently they check their email). In our evaluation, 10\% of users have the ability of
instant classification and the average delay of the normal classification is 2 hours~\cite{Mislove-NSDI-08}.

In \social, when a receiver classifies the email from a sender as spam, it issues
a behavioral report as \begin{center} ${[\rm spammer~report]}~I,~X\%$ \end{center}, 
where $I$ is the IP of the sender and $X$ is the confidence of this spammer report.
The issued spammer reports are gathered in the repository, and 
they are used when normal users with no capability of instant classification receive connections
from previously unencountered hosts. Each node has a view which is a subset of the \social\ network, 
and it only considers the spammer reports issued by nodes in its view. In addition, each view 
has pre-trusted users who have the capability of the instant classification, and the behavior reports issued by them are 
highly trustable. Therefore, classifier nodes shares their experiences by issuing spammer reports, and  normal nodes
use the reports to block spam from senders which they have not encountered before. In this evaluation, the size of the view is 
500 and the size of the pre-trusted set is 20.

The reporter trust that a \social\ node place on others is computed based on 
the direct trust value between each member of the view. This direct trust is in
turned computed based on the similarity of the spammer reports by the two members
of the view. Based on this direct trust value, each member of view gets the reporter trust value by using Dijkstra's algorithm.
Then, if the overall trust metric calculated by the equation \ref{eq:spammer-report-trust} is over 0.5,
a user blocks the SMTP connection.

\subsection{Spam Mitigation Effectiveness}

\begin{figure}[t]
\begin{center}
\subfigure[Percentage of blocked spam and legitimate email connections as a function of view size. Simulated time is 170h.]{\epsfig{file=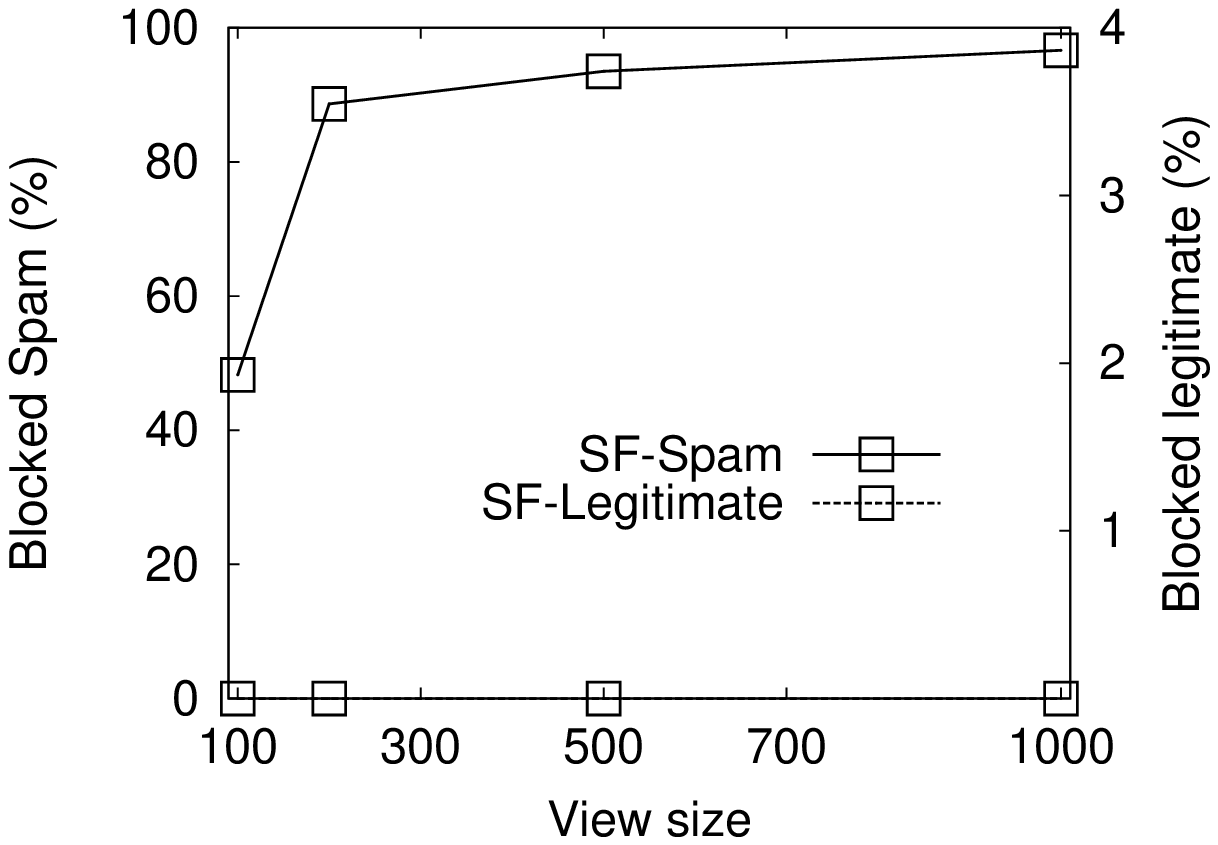, width = 0.24\textwidth}\label{fig:SFVariousView}}
\subfigure[Computation time of reporter trust as a function of the size of view.]{\epsfig{file=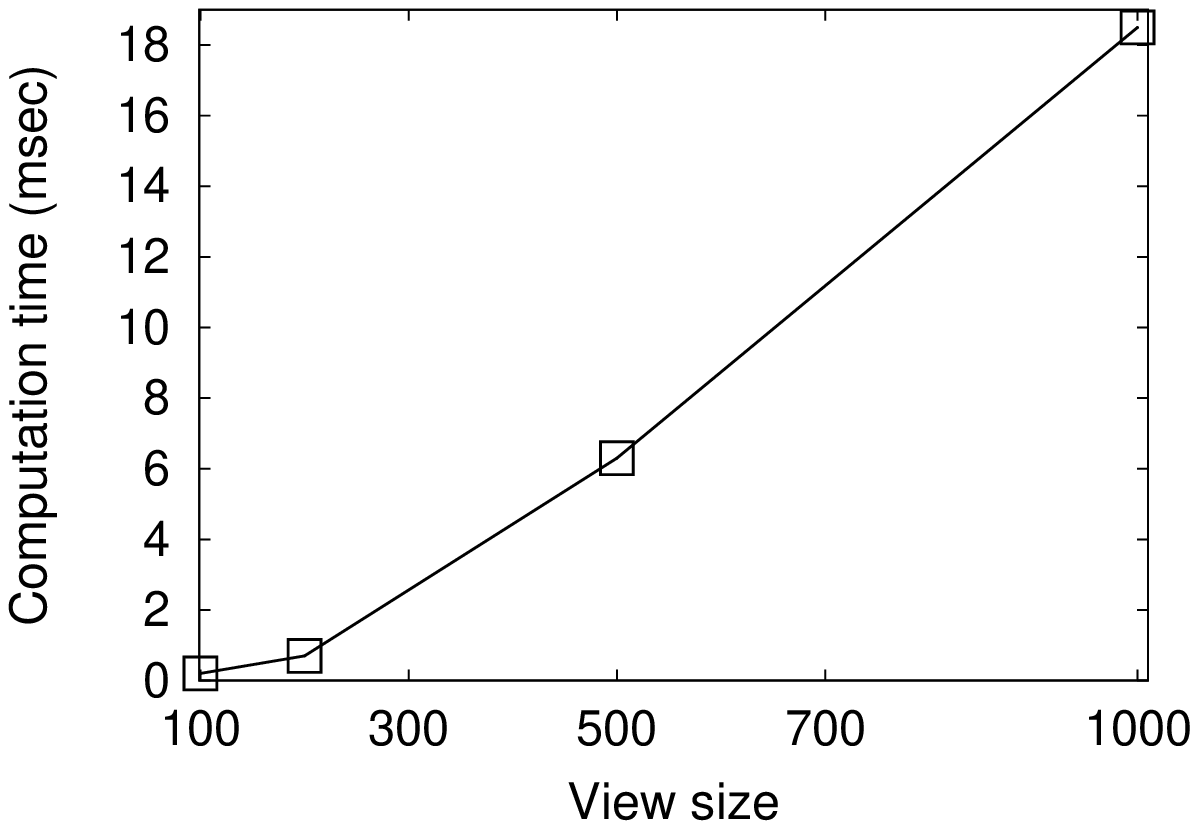, width = 0.24\textwidth}\label{fig:ReporterTrustComputation}}
\caption{Effectiveness of \social\ as a function of view size.}
\label{fig:VariousView}
\end{center}
\end{figure}

Before comparing \social\ and Ostra, we investigate the effectiveness of \social\ according to the size of view.
Figure \ref{fig:SFVariousView} shows the spam mitigation capacity of \social\ as a function of the size of view
when the simulated time is 170 hours. When the size of view is 100, \social\ blocks only around 50\% of spam email
connections.  On the other hand, when the size of view is 1000, \social\ blocks about 95\% spam emails.
Because spammers send spam emails to random targets, as the size of a node's view increases, the probability that a member
in the view encounters the spammer before the spammer contacts the node increases.
Additionally, once a user can get the spammer report from members of its view, 
it can block all the further spam emails from the detected spammers.

In order for a node to compute trustworthiness of email senders by using spammer reports,
it requires the reporter trust for all the members in its view.
The reporter trust is computed each time a node checks the likelihood of an email sender being spammer.
The reporter trust is computed by using Dijkstra's algorithm and the complexity of this computation on direct trust-annotated 
graph $T(V,E)$ is $O(|V|^2 log |V|)$. In Figure \ref{fig:ReporterTrustComputation}, we show the computation time of the reporter trust
with varying view size.  For the measurement, we use an Intel Core Duo P8600, 2.4GHz CPU, 3MB L2 cache, 4GB RAM machine,
and use reporter trust computation code written in Python 2.5.2. As the size of view increases, the computation time 
increase significantly, even though the performance of spam mitigation has already saturated with small size of view.
This result justifies our design choice to perform the reporter trust computation at the nodes and not the centralized
repository. Because the reporter trust is computed each time a node checks the trustworthiness of an email sender, 
a too large view size may becomes a performance bottleneck for nodes. Based on this results, 
we use 500 as the size of view hereafter.

\begin{figure}[t]
\begin{center}
\subfigure[Percentage of blocked spam and legitimate email connections as a function of \% of nodes being spammers. Simulated time is 170h.]{\epsfig{file=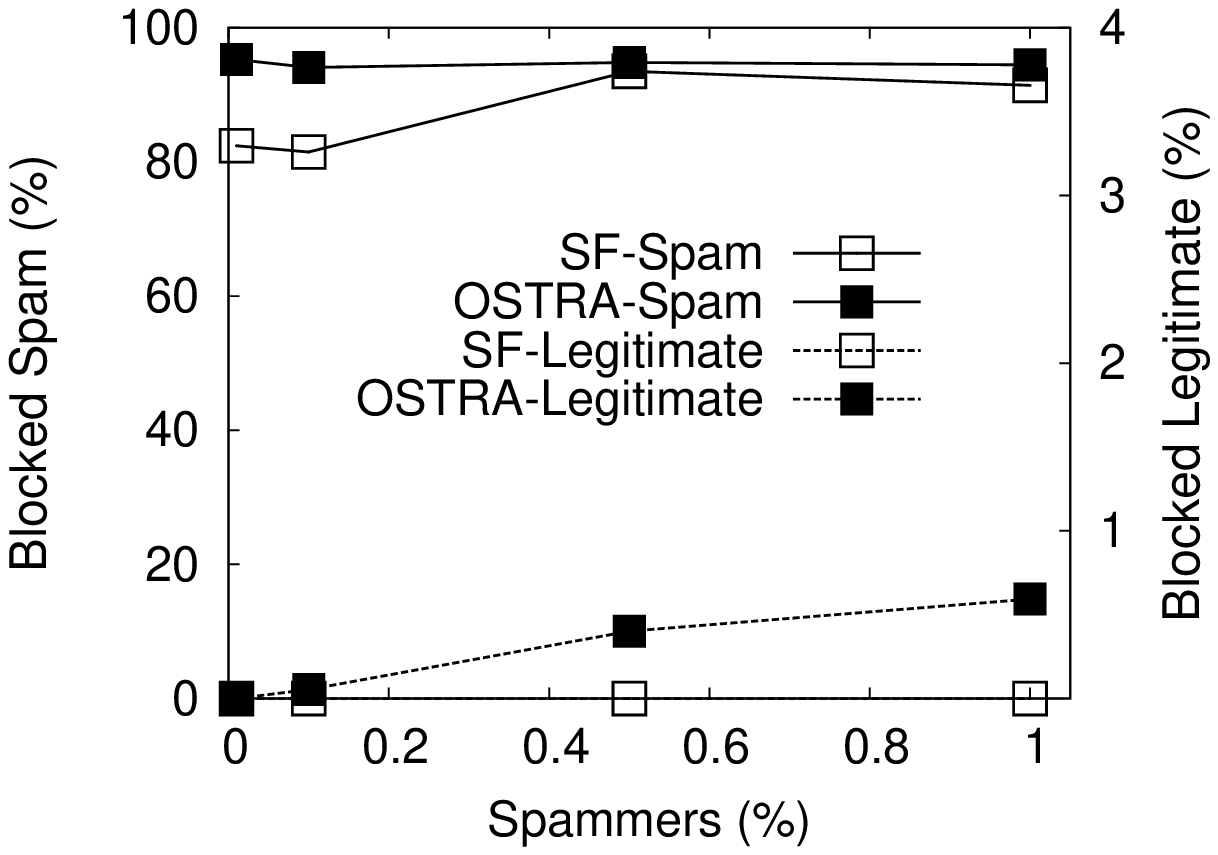, width = 0.24\textwidth}\label{fig:VariousSpammer}}
\subfigure[Percentage of blocked spam and legitimate emails connections as a function of simulated time. 0.5\% of nodes are spammers.]{\epsfig{file=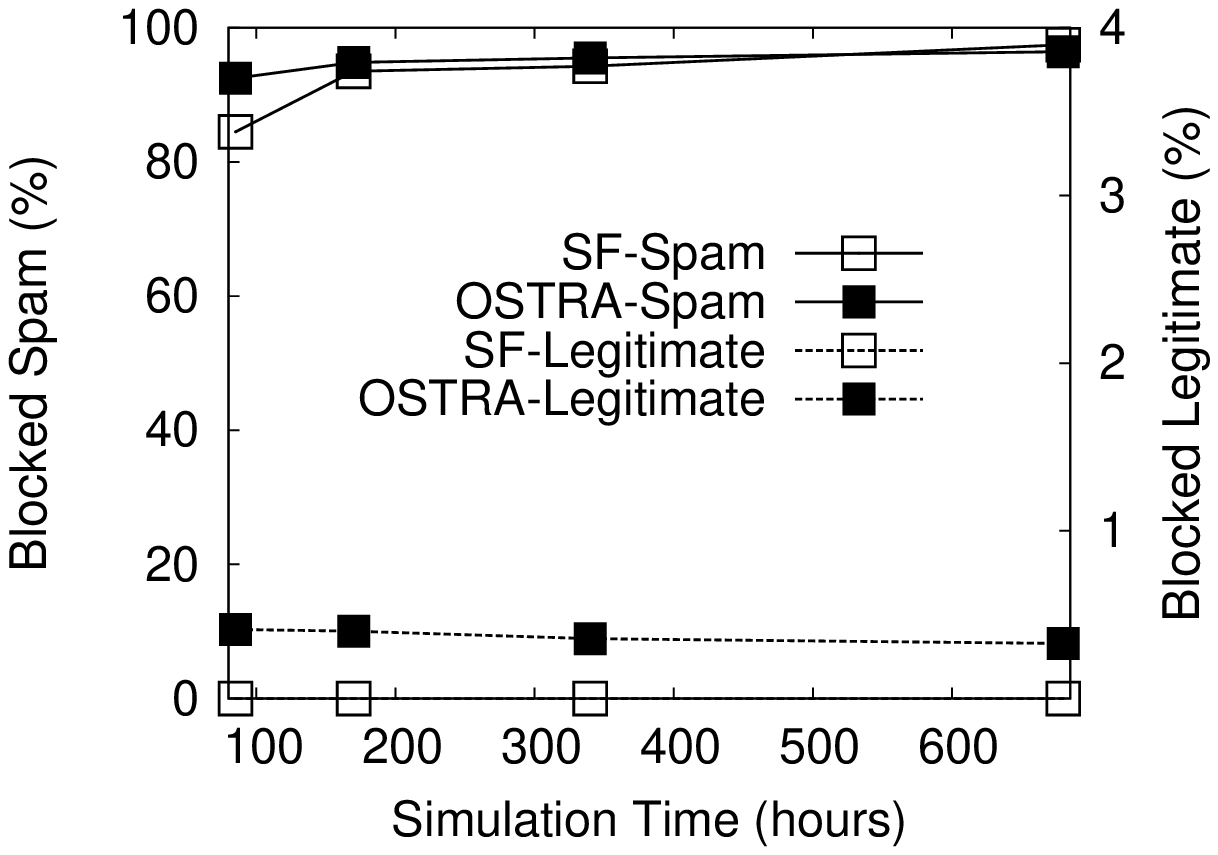, width = 0.24\textwidth}\label{fig:VariousTime}}
\caption{Effectiveness comparison of \social\ and Ostra.}
\label{fig:ComparisonSFvsOS}
\end{center}
\end{figure}

Figure \ref{fig:VariousSpammer} presents the spam mitigation effectiveness comparison between \social\
and Ostra under a varying number of spammers. We make two notes. The first one is that Ostra suffers from a non-negligible
false positive rate, and the second one is that \social\ allows more spam emails when there are small number of spammers.

In \social, a node only blocks an email sender only if it has been explicitly reported as spammer by a member of its view.
In these results, we assume that there is no false reporter, and there is no fake spammer report incriminating
legitimate users. This is the reason why \social\ does not suffer from false positive.
On the other hand, Ostra blocks all the links on the path used by a spammer, and some legitimate users can not send 
email because there is no available path in the social network. When the number of spammers are 1\% of nodes, which is 
500 spammers, around 0.6\% of total normal emails, which are about 6000 legitimate emails, can not be sent.

Ostra blocks about 94\% of spam connections regardless of the number of spammers.
Even though \social\ performs well for large number of spammers, it only blocks about 80\% of spam  when
the number of spammers is small, e.g., 0.1\% or 0.01\%. The main reason is the pre-trusted users which are included in every view.
Once, a pre-trusted node detects a spammer, every user can share the spammer report generated by it.
As the number of spammers increases, the probability that pre-trusted users early detect some of them increases.
Because of this early detection, \social\ can block more spam emails when the number of spammers are bigger.
Conversely, when the number of spammers is smaller, it is hard to detect them early, and \social\ allows proportionally
more spam emails.

\social\ requires some time for a node to detect spammers and during that time it allows
some spam connections to go through. But, once a user detects spammers via either referring to its view 
or classifying email senders itself, it will not allow any further spam emails.
To illustrate this characteristic, we show the performance with varying lengths of simulated time in Figure \ref{fig:VariousTime}.
In Ostra, after the spam blocking ratio becomes around 94\%, it will not change any more.
On the other hand, in SocialFilter, despite the spam blocking ratio being only 85\% at 85 hours, it increases
along with the simulated time and finally it blocks around 96\% spam email on 680 hours.
Eventually, unlike Ostra which suffers from the false positive as well as allows a portion of spam, 
almost all nodes in \social\ can block all the spam emails without any false positives.

\subsection{\social's Resilience to Attacks}
\label{subsec:socialite-resilience-to-attacks}

\begin{figure}[t]
\begin{center}
\subfigure[False Spammer Reports]{\epsfig{file=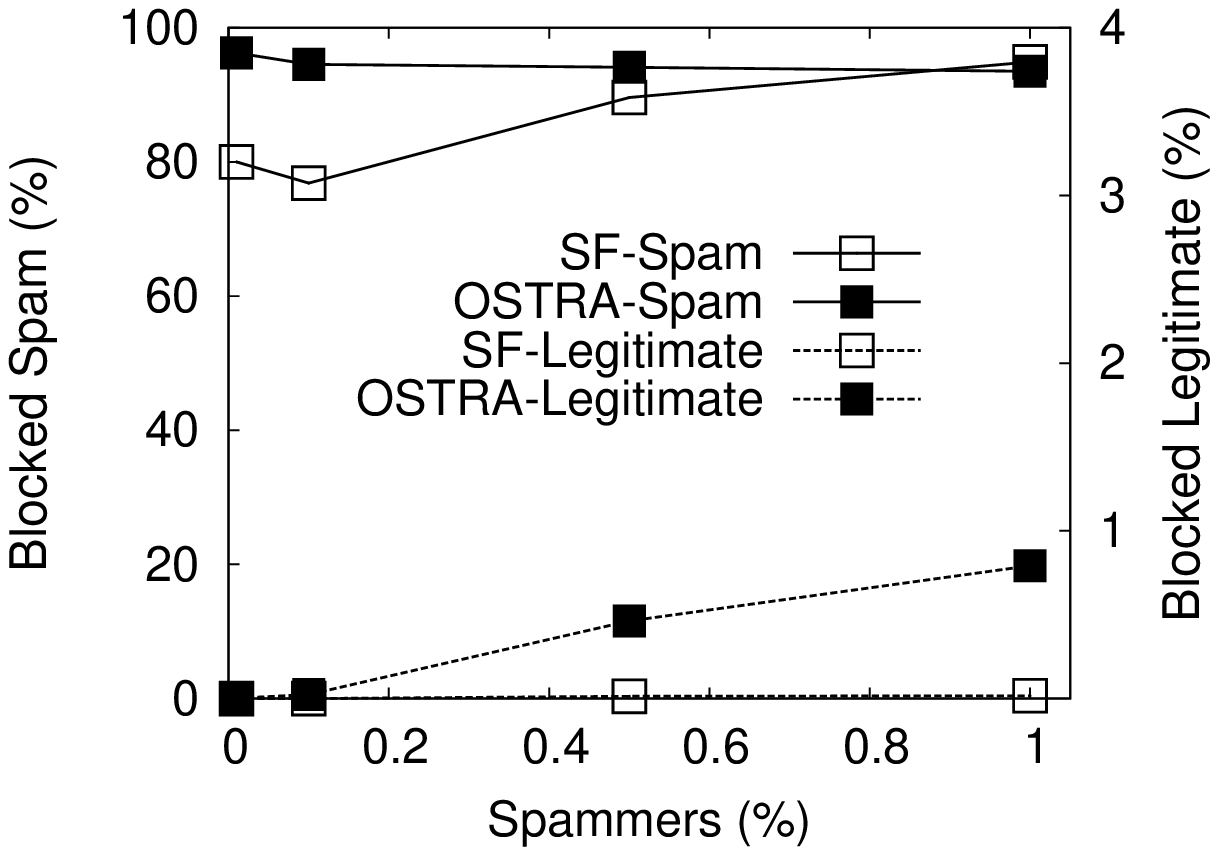, width = 0.24\textwidth}\label{fig:VariousFalseReporters}}
\subfigure[Sybil Attack]{\epsfig{file=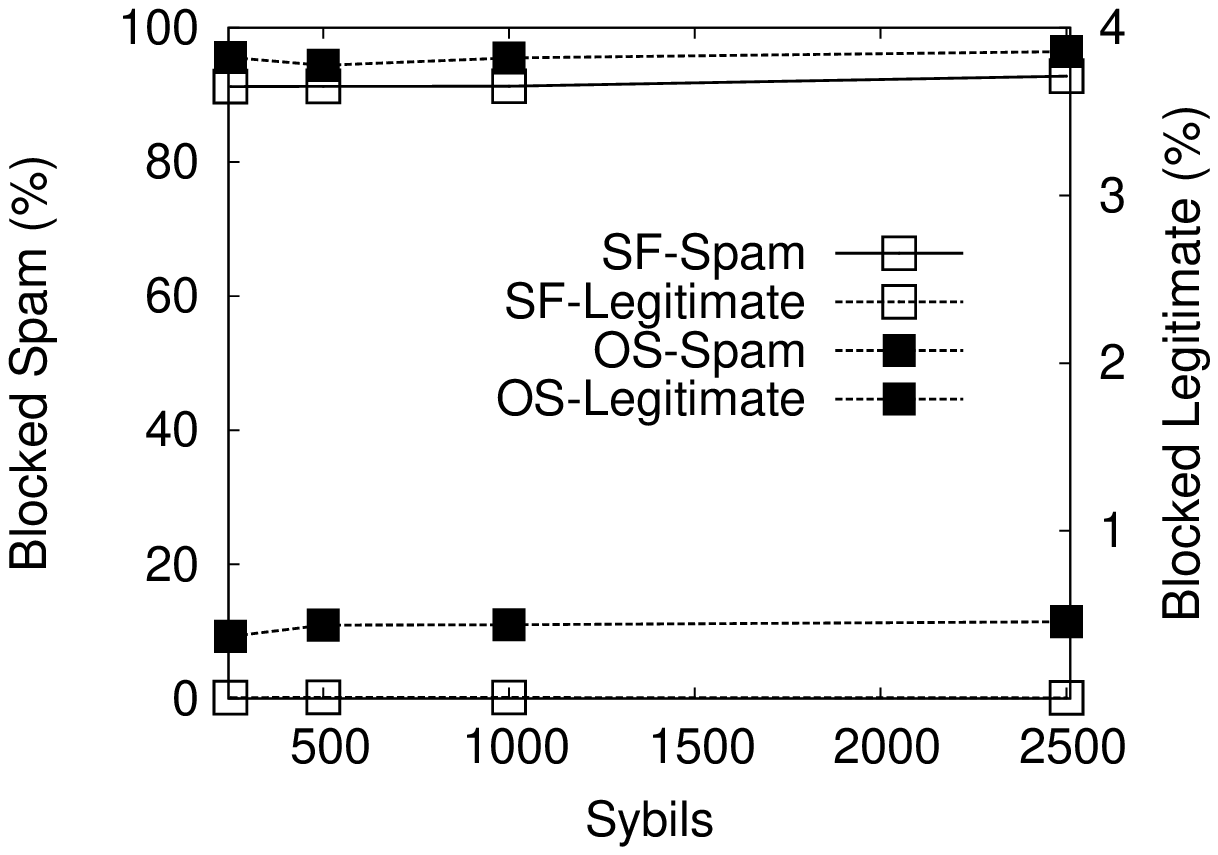, width = 0.24\textwidth}\label{fig:VariousSybils}}
\caption{Effectiveness comparison of \social\ and Ostra under attacks. 0.5\% nodes are spammers. Simulated time is 170h.}
\label{fig:UnderThreats}
\end{center}
\end{figure}

Spammers can collude for their spam emails to evade \social. First each spammer may not issue reports about colluding spammers.
Also each spammer falsely generates behavior reports about legitimate users to induce \social\ to block their emails.
To cheat Ostra, each spammer classifies a legitimate email and a spam email as unwanted email and legitimate, respectively.
Figure \ref{fig:VariousFalseReporters} shows the effectiveness of \social\ and Ostra as a function of the number 
of colluding spammers that reports falsely.

Although Ostra achieves the same effectiveness  in blocking spam connections as in the absence of
false reporters, the ratio of the false positives increases. Since Ostra does not have any 
method to recognize false classifications, Ostra is more adversely affected by the false reports.
On the other hand, \social\ achieves similar performance of blocking spam emails, and
the false positive rate due to false reports is very limited. 
This is because false reporters obtain very low direct trust to other legitimate users as their
spammer reports are different to reports of other legitimate users.
Eventually, the spammer reports issued by the false reporters are mostly ignored by
legitimate users.

Spammers may also create Sybil nodes to attack \social. However, in \social, Sybil users gets
very low identity uniqueness, which becomes even lower as the number of Sybil users increases.
Thus, despite spammers using Sybils that report falsely as well as send spam emails,
\social\ is resilient to this attack. We performed simulation with various number of Sybil users
and derived that that the performance of of the system is not substantially affected negatively (Figure \ref{fig:VariousSybils}).
In Ostra, Sybil spammers are blocked easily because the social links from the Sybils' creators 
have already been blocked. Although Ostra can block more spam emails, Ostra still suffers from false positives.

\section{Related Work}
\label{sec:related-work}

We now discuss prior work that is pertinent to \social's design
and is not discussed in the main body of the paper.

\social\ is inspired by prior work on reputation 
and trust management systems~\cite{Marti-ComNet-06,Hoffman-ACM-07,Blaze-SP-96}.
Well-known trust and reputation management systems include the rating 
scheme used by the eBay on-line auction site, object reputation systems
for P2P file sharing networks~\cite{Walsh-NSDI-06,Kamvar-WWW-03} and
PageRank~\cite{Brin-ComNet-98}.
In contrary to the above systems, our system incorporates social trust to mitigate false reporting 
and Sybil attacks. In addition, \social's view-based reporter trust scales better
than eigenvector-based trust metrics such as EigenTrust~\cite{Kamvar-WWW-03}, 
PageRank~\cite{Brin-ComNet-98} and TrustRank~\cite{Gyongyi-VLDB-04} because direct trust values between nodes 
change frequently and it would be expensive to consider the complete trust graph. In addition, 
eigenvector-based trust metrics do not provide an explicit confidence metric for a node, but
they only allow ranking the nodes instead.

\social\ is simular to SpamHaus~\cite{SpamHaus}, DShield~\cite{DShield} and TrustedSource~\cite{TrustedSource} 
in that it has a centralized repository. It differs in that it automates the process of evaluating reports
and assigning reputations to reporters. Thus it does not incur the management overhead
of traditional services, and can therefore scale to millions of reporters.

Prior work also includes proposals for collaborative spam
filtering~\cite{Zhong-INFOCOM-08,Zhou-Middleware-03,DCC}.   
CloudMark~\cite{CloudMark}, as does \social,
explicitly addresses the issue of trustworthiness of the collaborating spam reporters through a distributed reporter 
reputation management system based on history of past interactions. 
However, they do not leverage the social network to derive trust information.
Kong et al.~\cite{Kong-Computer-06} also consider untrustworthy reporters, 
using Eigentrust for deriving their reputation.  The aforementioned 
solutions only enable classifying the contents of
emails and not the source of spam. This requires email servers to waste
resources on email reception and filtering.  \social\ can assign trust 
metrics to sources, thereby rejecting unwanted email traffic on the outset.

Similar to \social, RepuScore~\cite{Singaraju-LISA-07} is also a collaborative reputation management
framework, which allows participating organizations
to establish sender accountability on the basis of senders
past actions. However, it does not exploit the social network of RepuScore server admins.

\social's identity uniqueness is based on SybilGuard and SybilLimit~\cite{Yu-SIGCOMM-06,Yu-SP-08},
where  the resource test in question is a  Sybil attacker's ability to create 
and sustain social acquaintances.  SybilGuard/Limit were designed to operate in 
a decentralized setting in which nodes are not aware of the complete 
social graph. We use a stripped-down centralized version of SybilLimit, 
because in our setting the OSN provider has complete knowledge of the social
graph's topology.

Prior work has also exploited trust in social networks to reliably assess the trustworthiness of entities
~\cite{Sabater-AAMAS-02,Pujol-AAMAS-02,Garris-NSDI-06,Zimmerman-95,Hogg-EC-04}. 
Unlike \social, they do not use social links to both bootstrap trust values 
between socially acquainted nodes and defend against Sybil attacks. 

\section{Conclusion}

We have presented \social, a large scale distributed system
for the rapid propagation of reports concerning the behavior 
of email senders. \social\ nodes use each other's reports and the social network 
of their human users to provide to applications a quantitative 
measure of an email sender's trustworthiness: the likelihood that the sender 
is spamming. Applications can in turn use this measure to make informed decisions 
on how to handle traffic associated with the host in question.

Our simulation-based comparative evaluation 
demonstrated our design's potential for the suppression of spam email. 
\social\  was able to identify $92\%$ of spam connections with greater
than $50\%$ confidence. Furthermore, in contrast to a competing social-network-based
spam mitigation technique, Ostra~\cite{Mislove-NSDI-08}, \social\ exhibited 
no false positives.

\let\oldthebibliography=\thebibliography
\let\endoldthebibliography=\endthebibliography
\renewenvironment{thebibliography}[1]{%
    \begin{oldthebibliography}{#1}%
    \setlength{\parskip}{0ex}%
    \setlength{\itemsep}{0ex}%
}{\end{oldthebibliography}}

{\scriptsize \bibliographystyle{abbrv} \bibliography{bibtex/facetrust,bibtex/dandelion}}

\end{document}